# Highly efficient nonvolatile magnetization switching and multi-level states by current in single van der Waals topological ferromagnet $Fe_3GeTe_2$

*Kaixuan Zhang*, Youjin Lee, Matthew J. Coak, Junghyun Kim, Suhan Son, Inho Hwang, Dong-Su Ko, Youngtek Oh, Insu Jeon, Dohun Kim, Changgan Zeng, Hyun-Woo Lee, and Je-Geun Park**

Dr. Kaixuan Zhang, Youjin Lee, Junghyun Kim, Suhan Son, Inho Hwang, Prof. Je-Geun Park
Center for Quantum Materials, Seoul National University, Seoul 08826, South Korea
E-mail: zaal@mail.ustc.edu.cn; jgpark10@snu.ac.kr

Dr. Kaixuan Zhang, Youjin Lee, Dr. Matthew J. Coak, Junghyun Kim, Suhan Son, Inho Hwang, Prof. Dohun Kim, Prof. Je-Geun Park
Department of Physics and Astronomy, and Institute of Applied Physics, Seoul National University, Seoul 08826, South Korea

Dr. Kaixuan Zhang, Youjin Lee, Dr. Matthew J. Coak, Junghyun Kim, Suhan Son, Inho Hwang, Prof. Je-Geun Park
Center for Correlated Electron Systems, Institute for Basic Science, Seoul 08826, South Korea

Dr. Dong-Su Ko, Dr. Youngtek Oh, Dr. Insu Jeon,
Samsung Advanced Institute of Technology, Suwon 16678, South Korea

Prof. Changgan Zeng
International Center for Quantum Design of Functional Materials, Hefei National Laboratory for Physical Sciences at the Microscale, CAS Key Laboratory of Strongly Coupled Quantum Matter Physics, Department of Physics, and Synergetic Innovation Center of Quantum Information & Quantum Physics, University of Science and Technology of China, Hefei, Anhui 230026, China

Prof. Hyun-Woo Lee
Department of Physics, Pohang University of Science and Technology, Pohang 37673, South Korea

Prof. Hyun-Woo Lee
Asia Pacific Center for Theoretical Physics, 77 Cheongam-ro, Nam-gu, Pohang 3773, South Korea

**Abstract:** Robust multi-level spin memory with the ability to write information electrically is a long-sought capability in spintronics, with great promise for applications. Here we achieve nonvolatile and highly energy-efficient magnetization switching in a single-material device formed of van-der-Waals topological ferromagnet $Fe_3GeTe_2$, whose magnetic information can be readily controlled by a tiny current. Furthermore, the switching current density and power dissipation are about 400 and 4000 times smaller than those of the existing spin-orbit-torque magnetic random access memory based on conventional magnet/heavy-metal systems. Most importantly, we also demonstrate multi-level states, switched by electrical current, which can dramatically enhance the information capacity density and reduce computing costs. Thus, our observations combine both high energy efficiency and large information capacity density in one device, showcasing the potential applications of the emerging field of van-der-Waals magnets in the field of spin memory and spintronics.

## 1. Introduction

Just over the past few years, van-der-Waals (vdW) magnetic materials have rapidly emerged as key members of the field of two-dimensional (2D) materials and device physics[1-6]. Due to their inherent superiorities like native-atomic thicknesses, high tunability, controllability, high compatibility, high designing or tailoring flexibility, to name only a few, they are tremendously promising for next-generation spintronic devices. Furthermore, the nm-thin features of magnetic vdW layers also facilitate compatibility with existing spintronic devices; besides, more opportunities will emerge for assembled heterostructures combined with a wide variety of other nonmagnetic vdW materials[7, 8]. Therefore, there is an exciting possibility that fully developed 2D vdW magnets will reshape the future of spintronics and computing science significantly[9-13].

$Fe_3GeTe_2$ (FGT) is at this stage the only reported topological vdW ferromagnetic metal. It exhibits well-behaved ferromagnet behavior with a large coercivity of several kOe at the nanoscale[14-16], naturally rendering it a robust magnetic memory prototype against external fluctuations. As we found in our previous work[17], the coercive field ($H_c$) of nm-thin FGT can be electrically reduced through the gigantic intrinsic spin-orbit-torque (SOT) mechanism[17-19]. This implies that we can readily control the magnetic information of such robust magnetic memory by applying electrical current, which is a core technique in spintronics[20].

Unlike conventional purely magnetic-field-controlled information switching, the current-controlled device promises much lower writing power consumption, larger memory density, and higher speed[21]. Such attractive characteristics have been the main driving force in the successful development of the Spin Transfer Torque Magnetic Random Access Memory (STT-MRAM), and also the newly emerging but extremely promising spin-orbit-torque-MRAM (SOT-MRAM)[22, 23]. Unfortunately, the present switching current density and power dissipation are still too high for conventional magnet/heavy-metal based SOT-MRAM[23], urging the development of new energy-efficient SOT-MRAM systems. In addition to these points, multi-bit magnetic memory is another crucial technique for future technology since it can significantly improve information density and reduce computing costs. The conventional ways to realize such multi-bit magnetic memory adopt individual memory cells in parallel or series[24-28]. In sharp contrast, if the multi-level memory cell can be realized in one device cell, the information capacity density and the work efficiency can be exponentially enhanced.

In this work, we succeed in achieving both high energy efficiency and large information capacity density in one single FGT device. The switching current density and power dissipation are suppressed by 400 and 4000 times compared with more conventional systems. Moreover, we can continually control the magnetization by current and obtain several states (about 8 states) with stability and nonvolatility. It enables approximately 3 bits rather than 1 bit, boosting the information density achieved and reducing computing costs. Such current-controlled nonvolatile multi-level states hold potential as multi-level spin memory, which is likely to be a critical component for the fascinating and rapidly-evolving field of spintronics.

## 2. Results and Discussion

### 2.1. Sample preparation and characterization

High-quality FGT single crystals were grown by the chemical vapor transport method and used for the device fabrication (See more details in the Methods section), following the technique detailed in our previous work[29]. As depicted in **Figure 1a**, the grown FGT crystals feature a layered hexagonal crystal structure with the space group *P*63/*mmc* (No. 194)[15, 30]. The as-grown FGT crystals are large, shiny, and well crystallized (Figure 1b). Figure 1c shows a high-angle annular dark-field scanning transmission electron microscopy (HAADF-STEM) image taken along the [120] zone axis. FGT hosts a vdW architecture with a vdW monolayer thickness and vdW gap of ~0.8 and ~0.3 nm, respectively[16]. The periodically ordered atomic arrangement is clearly outlined in the corresponding zoom-in image. We also carried out the HAADF-STEM and corresponding selected area electron diffraction (SAED) imaging along the [001] zone axis (Figure 1d), which match well with the hexagonal structure expected of FGT. The crystal plane indexes are indicated in the hexagonal SAED pattern, and the atomic arrangement also fits well in the zoom-in image. The in-plane lattice constants *a* and *b* and the interplanar spacing for the (100) plane are observed to be ~0.394 and ~0.34 nm, very close to reported values of 0.399 and 0.346 nm, respectively[30]. These characterizations demonstrate the high quality of our as-grown FGT crystals at both macroscopic and atomic scales. From the as-grown FGT crystals, we fabricated the FGT nanoflake devices (samples S1, S2, S3, and S4) with the Hall-bar-geometry electrodes. Figure 1e,f shows the typical optical image of an FGT device (sample S4) and the thickness of ~14.5 nm, confirmed by atomic force microscopy (AFM). The thickness of samples S1, S2, S3 are displayed in Figure S1.

### 2.2. Highly energy-efficient nonvolatile magnetization switching

As illustrated in **Figure 2a**, based on nm-thin metallic vdW ferromagnet FGT, we can readily realize a conceptual spin memory device, where magnetic information is written by current and read through the anomalous Hall effect, i.e., transverse Hall resistance ($R_{xy}$). Following conventional protocol, "0" and "1" states stand for the negative and positive saturated magnetization states. Figure 2b shows the $R_{xy}$-$H$ curves of an FGT device (sample S1) with various currents applied. As one can see, the coercivity is significantly reduced upon increasing current, consistent with the gigantic controlled coercivity by SOT in our previous work[17]. We note that one can change the information by measuring the hysteresis loop in our case with the conventional switching by the magnetic field (writing path I). Initially, the device is in the "0" state due to the spontaneous magnetization of FGT (green circle). Without writing current supplied (with a small $I_{\mathrm{write}}$=0.05 mA), the device exhibits a typical ferromagnet behavior, demonstrating a robust magnetic memory for information storage. By contrast, when a writing current is supplied ($I_{\mathrm{write}}$=2.0 mA), the coercive field is effectively reduced while the magnetization state is still well maintained. Therefore, one can readily control magnetic information with the assistance of current and magnetic field through the writing path I (indicated by the blue arrows in Figure 2b). First, we apply $I_{\mathrm{write}}$=2.0 mA, then increase the magnetic field from 0 to 0.1 T. As a consequence, one can switch a "0" state (green circle) to a "1" state (red circle) tracing the hysteresis loop. The reverse switching from "1" to "0" states can be realized equivalently.

Figure 2c demonstrates another novel magnetic information switching by directly controlling the writing current (writing path II; indicated by the blue arrows): After turning on a small fixed magnetic field, e.g., 0.1 T, one can apply $I_{\mathrm{write}}$ = 2 mA to the sample. As the coercivity can be significantly reduced by the current, the FGT device's magnetization will be switched by the current, from "0" to "1" state above a certain critical current. Figure 2d displays the corresponding switching process through the $R_{xy}$ ($H$)-$R_{xy}$ (-0.5 T) as a function of $I_{\mathrm{write}}$ (0 → 2 → -2 → 2 mA) under various magnetic fields from -0.5 T to 0.5 T. Initially, the FGT device is robustly in the "0" state ("-$M_s$" state) regardless of current sweep under -0.5 T. While ramping $I_{\mathrm{write}}$ from 0 to 2 mA under 0.1 T, the magnetization gradually increases to "+$M_s$" (indicated by the red arrow), i.e., the "1" state.

It is noteworthy that after reaching the "1" state, the magnetization is well maintained while sweeping current, demonstrating that such magnetic information switching from "0" to "1" states through $I_{write}$ is robust and nonvolatile. The reverse switching from "1" to "0" states is also realized via the same method. Additionally, due to the writing procedure's nonvolatility, the measured magnetization can readily read the information, say, $R_{xy}$, anytime via a small current without undesired writing.

Next, we show another device's systematic switching performance (sample S2), whose coercivity is also effectively reduced by the writing current (**Figure 3a**). Figure 3b exhibits the nonvolatile switching from "1" to "0" states by applying $I_{write}$ (maximum: 2 mA) under -0.06, -0.07, -0.08, and -0.09 T, and Figure 3c reveals that the system can also be reversely switched back from "0" to "1" state nonvolatile by using $I_{write}$ under 0.06, 0.07, 0.08, and 0.09 T. As expected, the critical switching current is larger under a smaller magnetic field.

In addition, we also demonstrate similar current-controlled switching behavior at higher temperatures of 100, 125, and 150 K, which are well above the liquid nitrogen temperature of 77 K (Figure S2). The device requires a well-defined ferromagnet behavior below its Curie temperature $T_c$; thereby, the device's working temperature is inevitably limited by the relatively low $T_c$ of ~200 K for FGT. It would be perfect if we could improve the operating temperature to room temperature. However, we can only secure the device's work temperature to be far above liquid nitrogen temperature at this stage. Although several ways were introduced to boost the $T_c$ of FGT, they are not compatible with current-controlled FGT-based SOT devices. For example, $T_c$ can be raised to room temperature for FGT by ionic gating[16]; however, it won't be realistic to be adopted for application with reactive ionic liquid and mA-order current. The most practical alternative for boosting $T_c$ can be carefully controlled doping into FGT while maintaining its topological bands and consequent large SOT. We tried to dope Co into FGT, but unfortunately, the Co-doped FGT was found to have even lower $T_c$ in our previous work[31]. It is still very challenging but worthy of seeking suitable methods or element doping for raising $T_c$ to room temperature while maintaining the band topology and large SOT of FGT.

For a memory device, the switching current density $J_{sw}$ and the switching power dissipation $J_{sw}^2/\sigma$ ($\sigma$ is conductivity) are two metrics used for energy efficiency. As summarized in Figure 3d, the switching current density and power dissipation across our FGT devices are on the order of $4\times10^5$ A/cm$^2$ and $2.5\times10^{14}$ W/m$^3$, much smaller than that of previous composite SOT systems. In particular, our $J_{sw}$ and $J_{sw}^2/\sigma$ are about 400 and 4000 times smaller than that of conventional heavy-metal/magnet SOT systems (For instance, $J_{sw}$ and $J_{sw}^2/\sigma$ of the Pt/Co system are ~$1.6\times10^8$ A/cm$^2$ and ~$1\times10^{18}$ W/m$^3$), highlighting the high energy efficiency of our FGT devices.

Due to different detailed SOT mechanisms, a direct comparison of current density between our work and conventional SOTs might not be perfectly rigorous. In particular, for our case, the SOT can directly reduce the perpendicular magnetic anisotropy or coercivity by the current. For more rigor, we make several notes here. Firstly, the magnitude of the applied magnetic field in our work is roughly similar to the other works in Figure 3d, which we used to visualize the high energy efficiency in our work. This is because the high energy efficiency comes from the unusual type of giant SOT in FGT and may also be assisted by an applied perpendicular magnetic field. Secondly, the conventional SOT systems function at room temperature. In contrast, the FGT-based SOT systems (including the previous Pt/FGT report and our work) only operate at lower temperatures below its $T_c$ ~ 200 K. Thereby, future optimizations are required, starting from such proof-of-principle demonstration of FGT-based SOT device.

In addition to room-temperature and high-efficiency, field-free switching is another long-sought technique for the current-driven device. To pursue field-free switching in our FGT system, the fundamental approach is replacing or removing the out-of-plane magnetic field via other means. Based on our knowledge and the achievements in the community of magnetic van-der-Waals materials, we have come up with two possible ideas:

One exciting possibility is combining the gating of (anti)ferromagnetic semiconductor and the proximity effect in an FGT/semiconductor heterostructure. The conceptual design is illustrated below: We can make FGT proximal to an out-of-plane (anti)ferromagnetic semiconductor like $CrI_3$, whose spins can be switched by an out-of-plane electrical field without energy dissipation[32]. In such a case, electrical-field gating can make adjacent (anti)ferromagnetic semiconductor $CrI_3$ change its magnetization configuration, giving an equivalent magnetic field to FGT. Therefore, the external magnetic field is not indispensable in this FGT/$CrI_3$ heterostructure. Combining with the SOT effect in FGT, we can, in principle, realize the field-free SOT-MRAM. However, this complicated work needs numerous experimental efforts and careful handling.

Except for the aforementioned exchange-biased field in the FGT/$CrI_3$ heterostructure, one can induce an additional effective out-of-plane field to FGT via lateral symmetry breaking. A previous work[33] demonstrated that an effective $H_z$ field could be induced in a system with lateral symmetry breaking through the wedge oxide film capping. More importantly, such $H_z$ field orientation and strength can be controlled by the current direction and amplitude. Therefore, if the same principle is applied to our FGT system with large intrinsic SOT, one can quickly realize the field-free switching. It would be very challenging but worthy of finding suitable ways to break the lateral symmetries of FGT in the future, such as the proper target oxide films for wedge capping.

### 2.3. Multi-level states by the current

We emphasize that intermediate magnetization states between "0" and "1" states can also be achieved by controlling the maximum of $I_{\mathrm{write}}$ or the magnetic field's magnitude. As demonstrated in **Figure 4a**,**b**, under a smaller magnetic field (±0.04 T and ±0.05 T) with a maximum current of $I_{\mathrm{write}}$=2 mA, the system enters a nonvolatile intermediate state. Such an intermediate state can be universally reproduced, as shown in Figure 4c for another device (sample S3). This behavior is nearly essential for a device to define several different step-like magnetization states between "0" and "1" states, which can be exploited as a multiple-state spin memory. Such multiple states enable several digitals (multi bits) more than two digitals (1 bit) to improve information density and reduce computing costs.

Next, we demonstrate such attractive multi-level states by simply controlling the writing current in **Figure 5**. Figure 5a exhibits the $R_{xy}$ ($H$)-$R_{xy}$ (-0.5 T) as a function of $I_{\mathrm{write}}$ (0 → 0.5 → -0.5 → 0.5 → 0 mA; then 0 → 0.75 → -0.75 → 0.75 → 0 mA) under -0.06 T of applied field. The system gradually enters different intermediate states between the "1" and "0" states. We extracted from Figure 5a the $R_{xy}$ ($H$)-$R_{xy}$ (-0.5 T) as a function of time with different maximum $I_{\mathrm{write}}$ values in Figure 5b. The blue curve represents the time sequence of $I_{\mathrm{write}}$, while the red curve denotes the corresponding time sequence of the magnetization. Multiple step-like states were achieved from the "1" to "0" states. To better reveal this point, we directly show the realized multiple steps controlled by the maximum value of $I_{write}$ in Figure 5c, where roughly 8 states exist, including the "0" and "1" states. Therefore, we can continually control the magnetic state by current and divide it into several digital steps with stability, featuring a promising platform for nonvolatile multi-level spin memory.

Here we would like to discuss the possible influence of Barkhausen effect on the multi-level states in our FGT devices. Figure S4 shows the current-controlled multi-level states as a function of time for much longer than 5 seconds, where the multi-level states are seen to be considerably stable. It indicates that the Barkhausen effect cannot be a minor effect in explaining our observations. On the other hand, a previous work reported that the domain size of an FGT nanoflake is larger than ~0.5 μm below 110 K[34]. The transverse Hall electrodes in our devices have a distance of about 1~2 μm, which covers only 2~3 FGT domains at most. Therefore, the Barkhausen effect on domain wall movement tends to be weak in our FGT device, which supports the stable multi-level states shown above.

The essence of the multi-level states reported here is that the magnetization can be continuously controlled by a current, which we believe is coming from the intrinsic SOT in FGT. Unlike the conventional SOT, the intrinsic SOT of FGT itself can be perfectly incorporated in its static-free energy, modifying the magnetic anisotropy by current[17, 18]. It indicates that the current-induced effective SOT field in FGT functions as an effective magnetic field that can align the spins to a certain direction with the external magnetic field, leading to the continuously controlled magnetism.

The highly efficient nonvolatile switching and multi-level states through current in FGT

devices can be further developed and exploited through the well-known three-terminal SOT-MRAM[24,25], where the writing and reading are physically decoupled due to independent current paths, leading to much-improved reliability and optimization flexibility. In SOT-MRAM, information can be written by the current through the SOT effect but read through the tunneling magnetoresistance (TMR) effect of a magnetic tunnel junction (MTJ), rather than the AHE adopted here in our work. The passing writing current can continuously manipulate the tunneling resistance as the magnetization did, leading to a multi-level SOT-MRAM. Such multi-level SOT-MRAM has much better performance and would be a more practical alternative for spin memory and spintronics.

Although there are many requirements for a commercial, practical SOT memory, we cannot achieve every aspect in our proof-of-principle demonstration of this single work. Nonetheless, we hope the highly efficient nonvolatile switching, multi-level states, and possible field-free designs in this FGT-based SOT system can bring enough importance and inspiration to the field of magnetic van-der-Waals materials and spintronics.

## 3. Conclusion

In summary, we systematically investigated the magnetization switching behavior, controlled by current, of the magnetic van der Waals $Fe_3GeTe_2$. We achieved both nonvolatility and extremely high energy efficiency due to a small current density being required. Moreover, through continually altering the magnetization by controlling the writing current, we demonstrated multi-level states with several stable digital steps. Our work provides the proof-of-principle demonstration for a novel memory concept. These exciting properties can be easily incorporated into the promising SOT-MRAM architecture, bringing the new magnetic van-der-Waals materials toward applications in spin memory and spintronics.

## 4. Experimental Section

*Sample preparation:* High-quality FGT single crystals were synthesized by the chemical vapor transport method with iodine as the transport agent. High-purity elements (Fe, Ge, Te) were stoichiometrically mixed with iodine and sealed in a vacuum quartz tube. The tube was placed in a two-zone furnace to grow crystals with a temperature gradient of 750 ˚C (source) to 680 ˚C (sink) for 7 days.

FGT nanoflakes were exfoliated onto the $SiO_2$/Si substrate inside the Ar-filled glove box by the conventional mechanical exfoliation method from the as-grown single crystals. The samples were sealed in a vacuum plastic package inside the glove box and then taken out. Before the standard electron beam lithography (EBL), the spin-coated PMMA polymer on FGT was baked with a moderate condition: 130 ˚C for 1.5 min. Immediately after EBL, 75/5 nm Au/Ti electrodes were deposited by the electron beam evaporation under a high vacuum ($<10^{-5}$ Pa).

We note that the noise level of switching data is determined mainly by the device quality, particularly the contact resistance between electrodes and FGT. For example, the contact resistance is ~5 KΩ, ~1 KΩ, and ~0.1 KΩ for samples S2, S3, and S1, leading to a reasonable noise level of sample S2 > sample S3 > sample S1. Of course, future practical FGT-based devices need to establish a more stable and well-behaved fabrication procedure to minimize contact resistance.

Nevertheless, the stability and reproducibility of the reported switching behavior are not problematic in the abundant switching curves of our work.

*Electrical transport measurements:* We perform transport measurements using a homemade resistivity probe operated inside a Quantum Design physical property measurement system with a maximum magnetic field of 9 T.

**Supporting Information**

Supporting Information is available from the Wiley Online Library or the author.

**Acknowledgments**

We thank Prof. Cheol Seong Hwang and Nahyun Lee for their support and helpful discussions. CQM was supported by the Leading Researcher Program of the National Research Foundation of Korea (Grant No. 2020R1A3B2079375), and this work was partially supported by the Institute for Basic Science (IBS) in Korea (Grant No. IBS-R009-G1). The theoretical works at the POSTECH were funded by the National Research Foundation (NRF) of Korea (Grant No. 2020R1A2C2013484). In addition, the Samsung Advanced Institute of Technology also supported this work.

Received: ((will be filled in by the editorial staff))
Revised: ((will be filled in by the editorial staff))
Published online: ((will be filled in by the editorial staff))

**Figures**

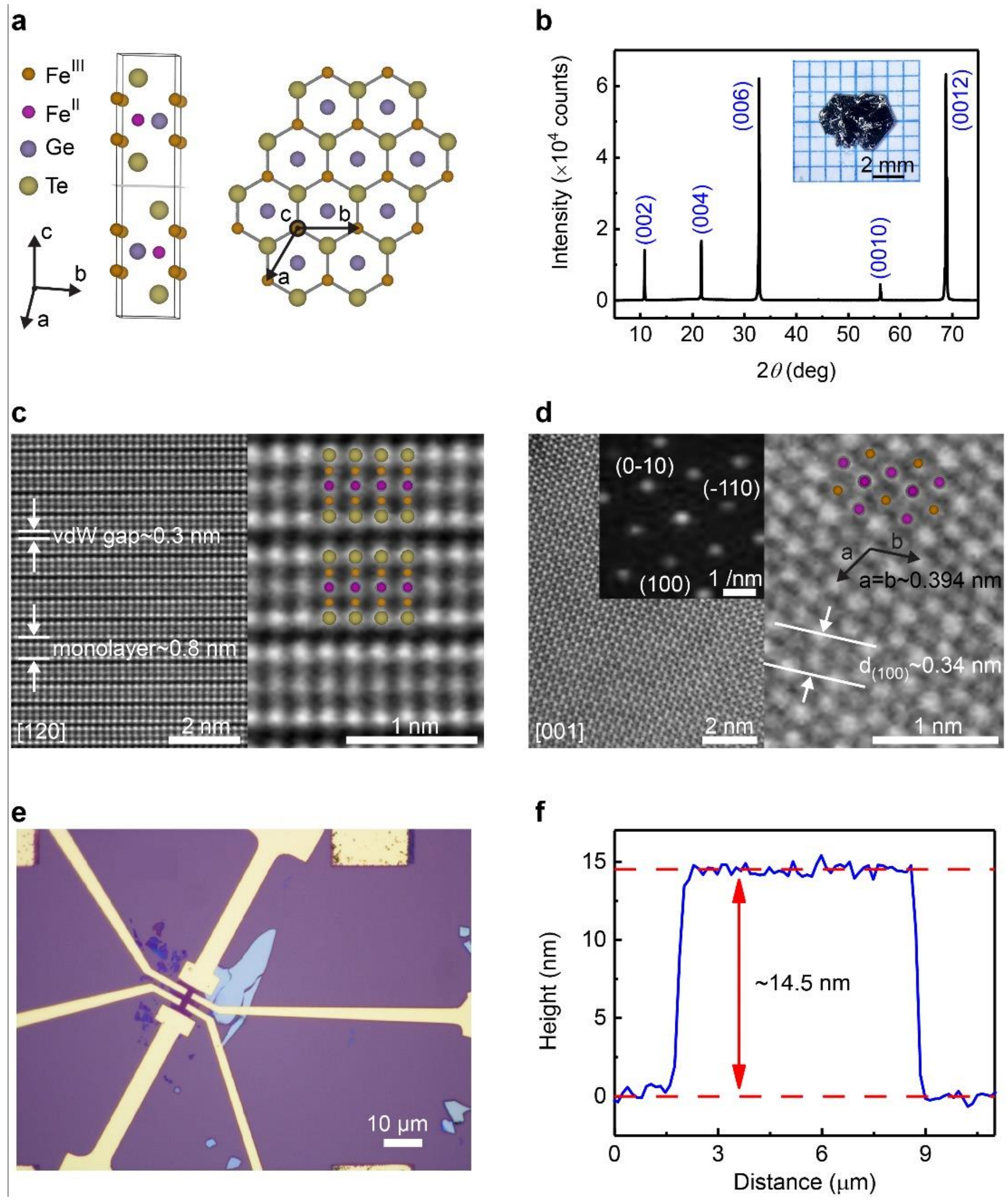

**Figure 1.** Structure and device characterizations. a) Crystallographic structure of $Fe_3GeTe_2$. $Fe^{III}$ and $Fe^{II}$ denote the two inequivalent Fe sites with oxidation numbers +3 and +2, respectively. The black lines indicate the unit cell. Right: the crystal structure of a monolayer viewing along the c axis. b) X-ray diffraction pattern of an FGT crystal. The inset is the optical image of the crystal. c) HADDF-STEM image of an FGT single crystal along the [120] direction, where the vdW layer and gap are marked. The scale bar is 2 nm. Right: zoom-in image, where the periodically ordered

atomic arrangement is outlined, and the scale bar is 1 nm. d) HAADF-STEM, and SAED images taken along the [001] direction. The crystal plane indices are marked, and the scale bars are 2 nm and 1 /nm, respectively. Right: zoomed-in image, where the in-plane lattice, interplanar spacing, and atomic arrangement are depicted. The scale bar is 1 nm. e) The optical image of a typical FGT device with the Hall-bar-geometry electrodes (sample S4). The scale bar is 10 μm. f) The thickness of FGT is ~14.5 nm, measured by the AFM.

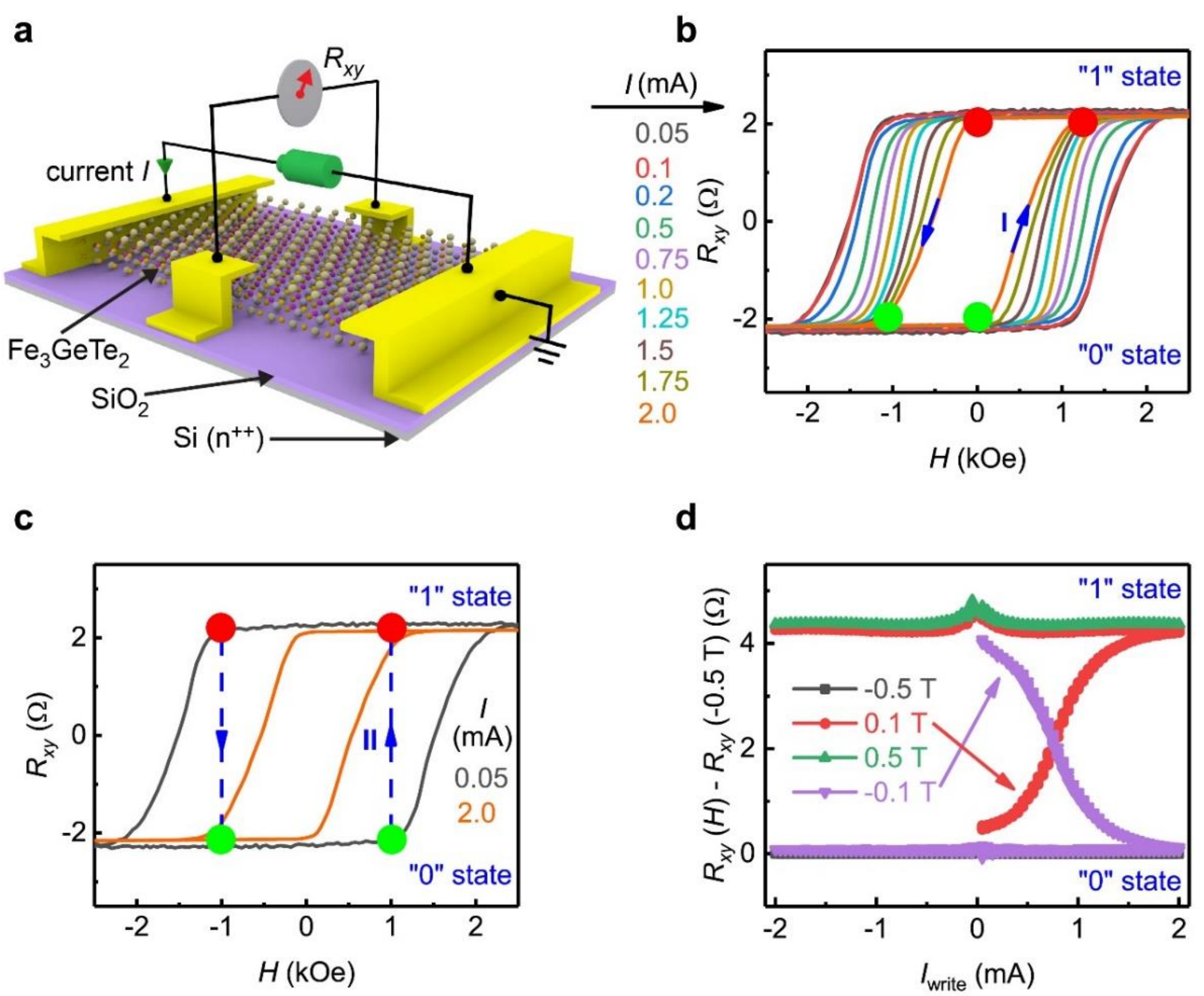

**Figure 2.** Nonvolatile magnetization switching. a) Schematic of a conceptual spin memory device based on FGT. The magnetic information is written by the current $I_{write}$ while read by transverse

Hall resistance $R_{xy}$. b) $R_{xy}$-$H$ curves of sample S1 at 2 K with various currents. The blue arrows illustrate the conventional information switching by magnetic field for writing path I. c) $R_{xy}$-$H$ curves of sample S1 with applied current $I = 0.05$ and 2 mA. This illustrates the transition between "0" and "1" states by the writing current through a novel-writing path II (indicated by the blue arrows). d) $R_{xy}(H)$-$R_{xy}$ (-0.5 T) as a function of $I_{\mathrm{write}}$ ($0 \rightarrow 2 \rightarrow -2 \rightarrow 2$ mA) under various magnetic fields from -0.5 T to 0.5 T. The red and purple arrows indicate the switching from "0" to "1" and "1" to"0" states while ramping $I_{\mathrm{write}}$ from 0 to 2 mA, respectively.

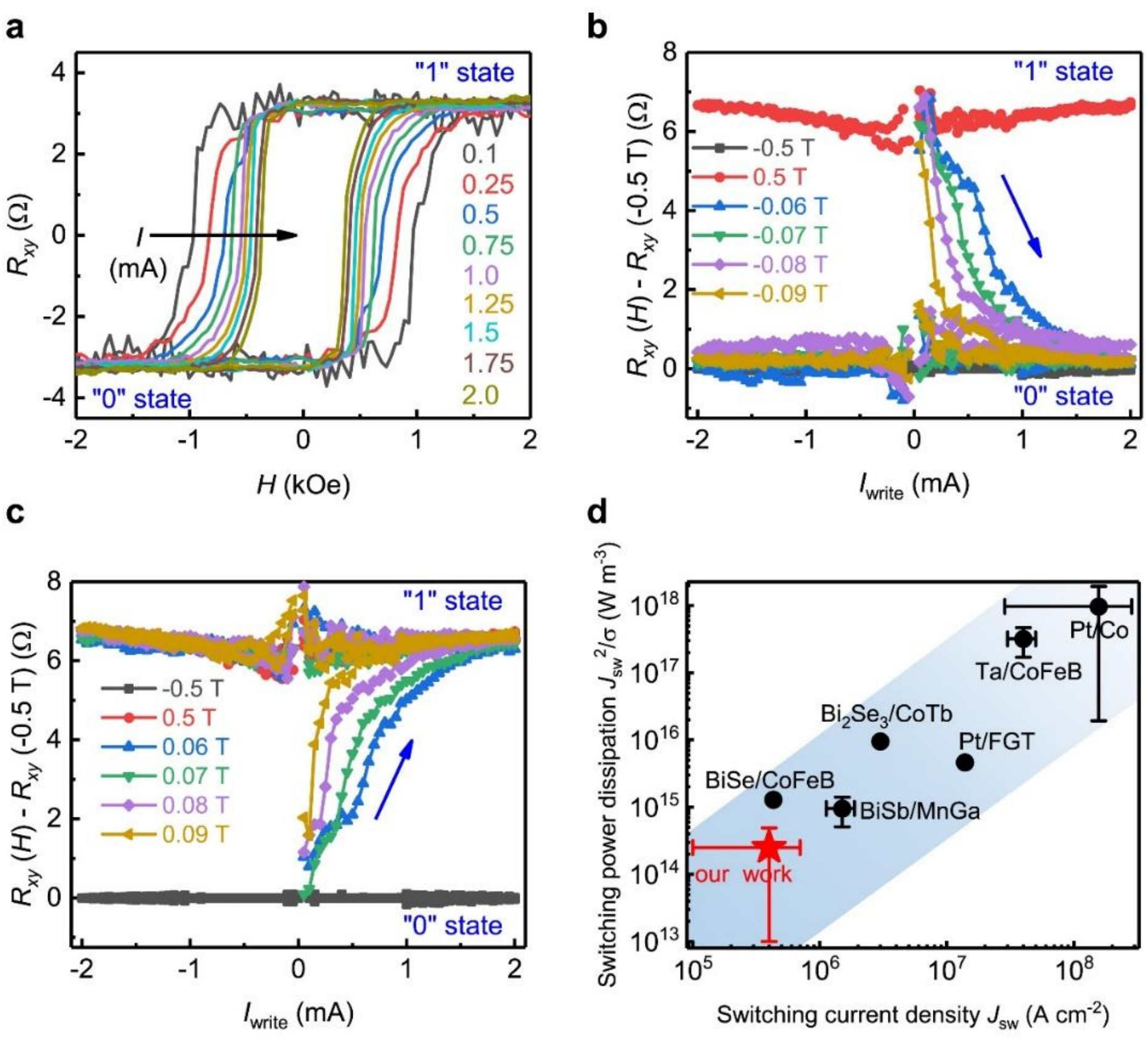

**Figure 3.** Highly energy-efficient nonvolatile magnetization switching by current with systematic

performance. a) $R_{xy}$-$H$ curves of sample S2 at 2 K with various current $I$. b) $R_{xy}$ ($H$)-$R_{xy}$ (-0.5 T) as a function of $I_{\text{write}}$ (0 → 2 → -2 → 2 mA) under various magnetic fields. The blue arrow indicates the transition from "1" to "0" states while ramping $I_{\text{write}}$ from 0 to 2 mA under -0.06, -0.07, -0.08, and -0.09 T. c) $R_{xy}$ ($H$)-$R_{xy}$ (-0.5 T) as a function of $I_{\text{write}}$ (0 → 2 → -2 → 2 mA) under various magnetic fields. The blue arrow indicates the transition from "0" to "1" states while ramping $I_{\text{write}}$ from 0 to 2 mA under 0.06, 0.07, 0.08, and 0.09 T. d) The switching current density $J_{\text{sw}}$ and switching power dissipation $J_{\text{sw}}^2/\sigma$ ($\sigma$ is conductivity) of our FGT devices and various composite SOT systems[23, 35-38]. The switching process of the Pt/Co and Ta/CoFeB systems relates to the single domain state while other systems involve the multi-domain states. The device works at lower temperatures below $T_c$~ 200 K for the Pt/FGT report and our work, while others were performed at room temperatures.

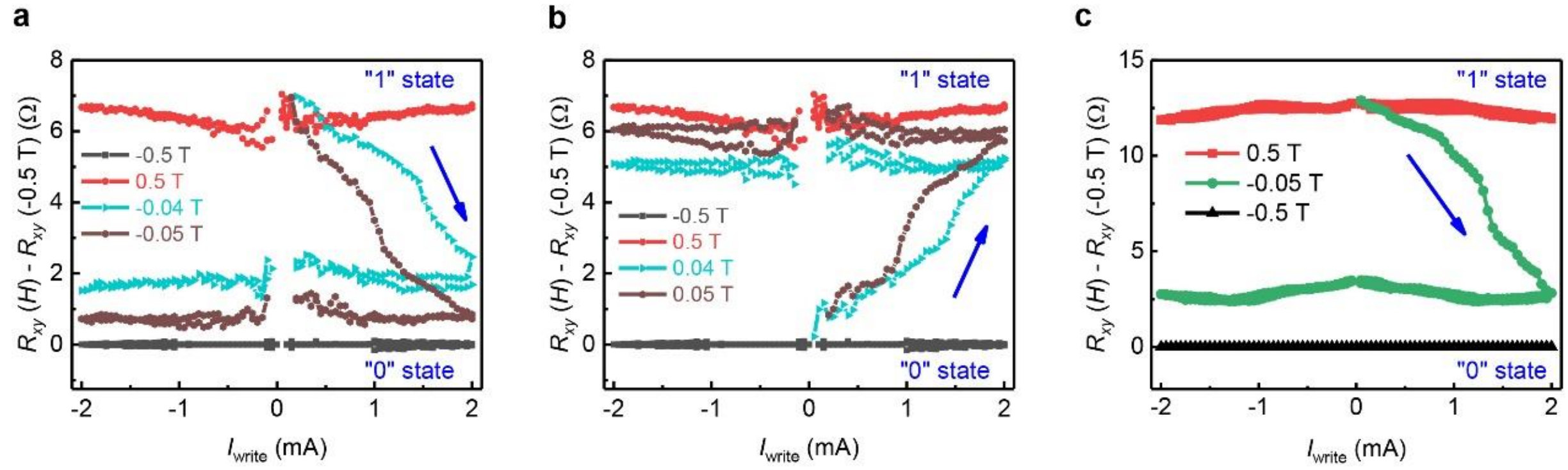

**Figure 4.** Achieving intermediate states between "0" and "1" states. a) $R_{xy}$ ($H$)-$R_{xy}$ (-0.5 T) of sample S2 as a function of $I_{\text{write}}$ (0 → 2 → -2 → 2 mA) under various magnetic fields. The blue arrow indicates the transition from "1" to an intermediate state close to the "0" state while ramping

$I_{write}$ from 0 to 2 mA under -0.04 and -0.05 T. b) $R_{xy}$ ($H$)-$R_{xy}$ (-0.5 T) of sample S2 as a function of $I_{write}$ (0~2~-2~2 mA) under various magnetic fields. The blue arrow indicates the transition from "0" to an intermediate state close to "1" state while ramping $I_{write}$ from 0 to 2 mA under 0.04 and 0.05 T. c) $R_{xy}$ ($H$)-$R_{xy}$ (-0.5 T) of sample S3 as a function of $I_{write}$ (0 → 2 → -2 → 2 mA) under various magnetic fields. The blue arrow indicates the transition from "1" to an intermediate state while ramping $I_{write}$ from 0 to 2 mA under -0.05 T.

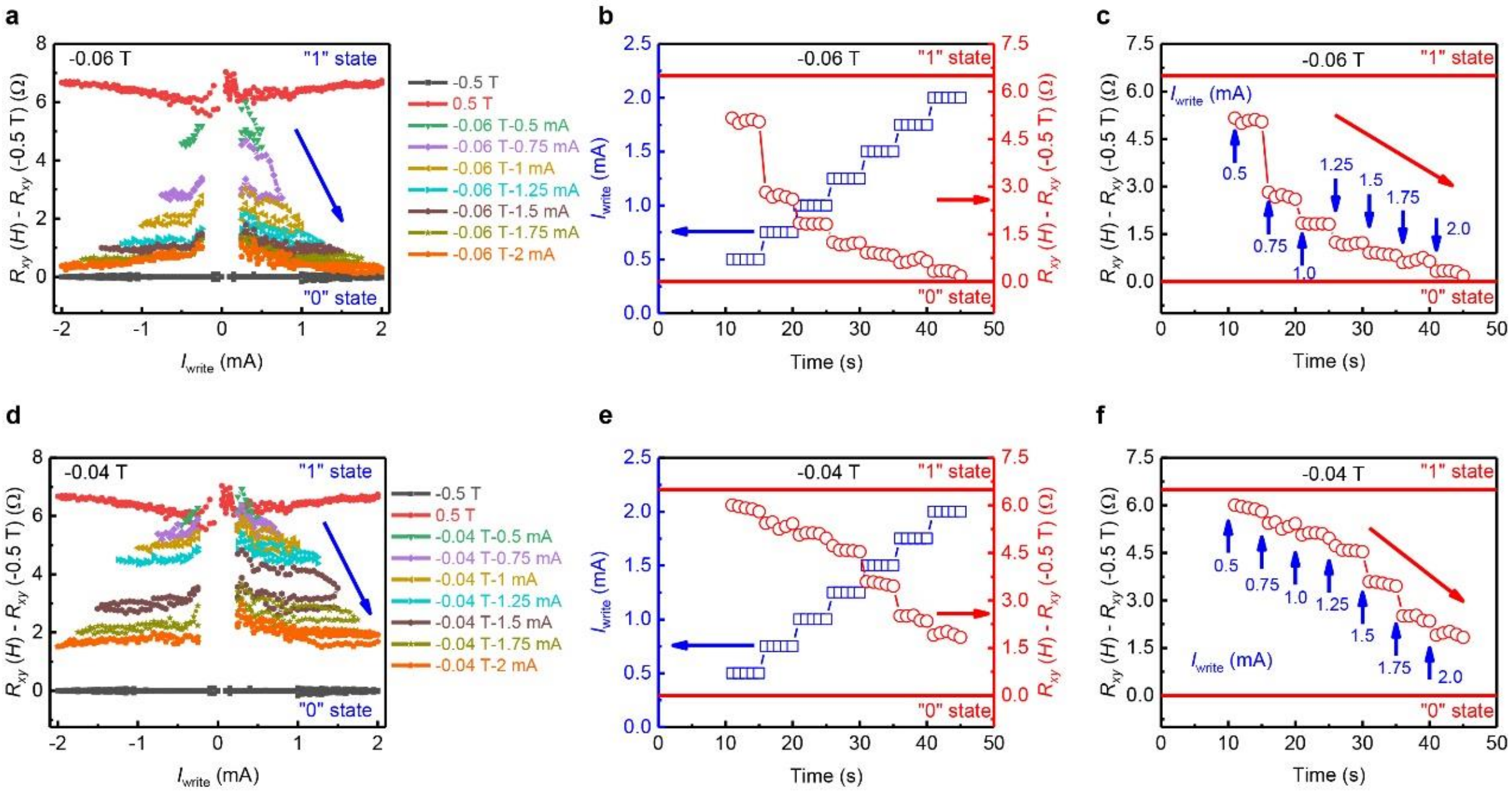

**Figure 5.** Multi-level states are achieved via electrical current with systematic performance. a) $R_{xy}$ ($H$)-$R_{xy}$ (-0.5 T) as a function of $I_{write}$ (0 → 0.5 → -0.5 → 0.5 → 0 mA; then 0 → 0.75 → -0.75 → 0.75 → 0 mA) under -0.06 T. The system gradually enters into different intermediate states from "1" to "0" states (indicated by the blue arrow). b) $R_{xy}$ ($H$)-$R_{xy}$ (-0.5 T) as a function of time with different maximum $I_{write}$. The blue curve represents the time sequence of $I_{write}$, while the red curve

denotes the magnetization's corresponding time sequence. Multiple step-like states were achieved from “1” to “0” states. c) $R_{xy}$ ($H$)-$R_{xy}$ (-0.5 T) as a function of time with different maximum $I_{write}$. The blue arrows denote the value of applied $I_{write}$ as time passed. Multiple step-like states were achieved from “1” to “0” states (indicated by the red arrow). d-f) The same behavior but under a magnetic field of -0.04 T.

## The table of contents entry

We achieved the nonvolatile and highly energy-efficient magnetization switching by current in single van der Waals topological ferromagnet $Fe_3GeTe_2$. The switching current density and power dissipation are about 400 and 4000 times smaller than conventional magnet/heavy-metal-based spin-orbit-torque magnetic random access memory. In addition, we further realize the multi-level states by the current, which can dramatically enhance the information capacity density and reduce computing costs. Our observations combine high energy efficiency and large information capacity density in one device, bringing the nascent van der Waals magnets toward spin memory and spintronics.

Kaixuan Zhang*, Youjin Lee, Matthew J. Coak, Junghyun Kim, Suhan Son, Inho Hwang, Dong-Su Ko, Youngtek Oh, Insu Jeon, Dohun Kim, Changgan Zeng, Hyun-Woo Lee, and Je-Geun Park*

Title: Highly efficient nonvolatile magnetization switching and multi-level states by current in single van der Waals topological ferromagnet $Fe_3GeTe_2$

ToC figure

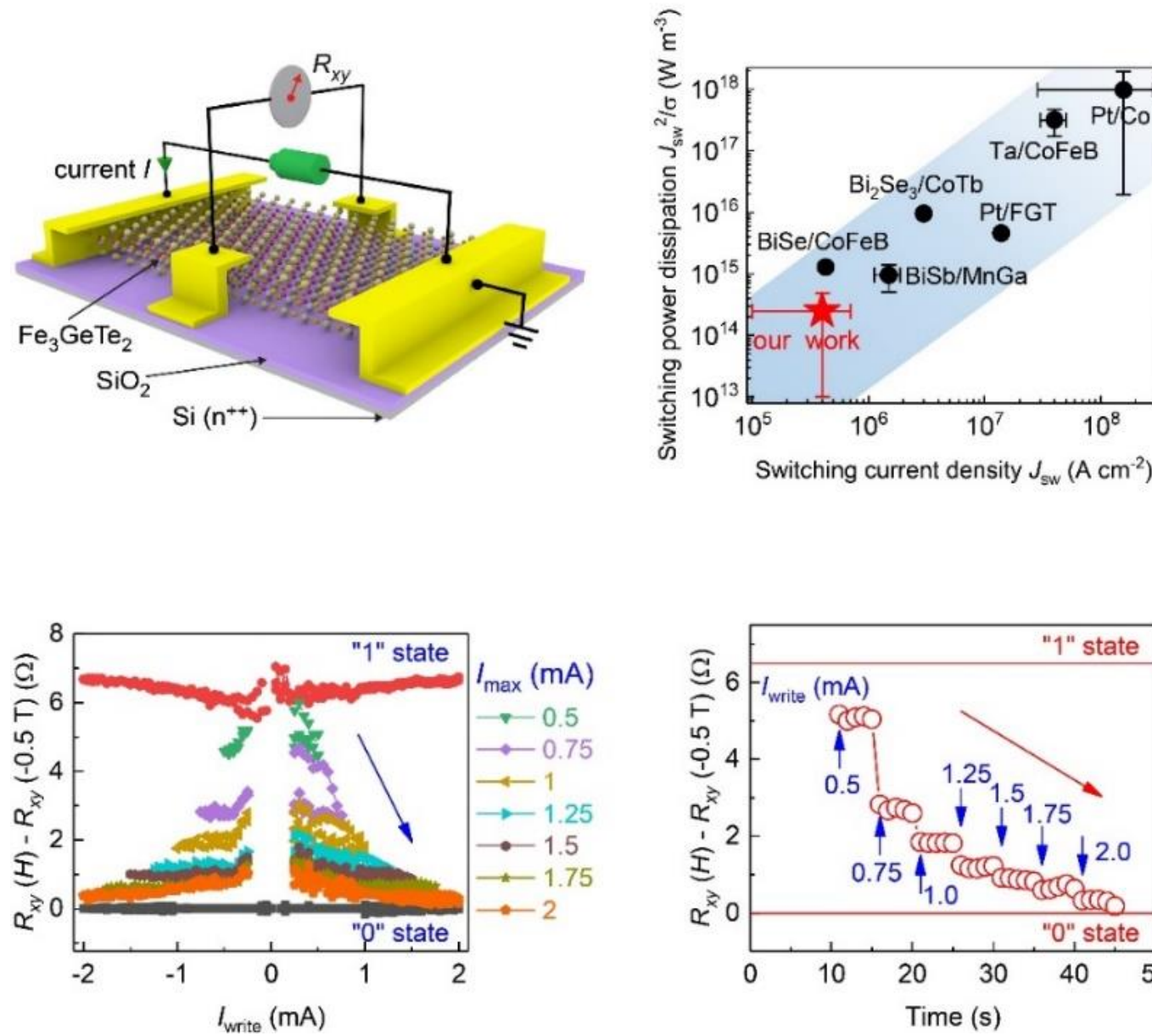

# Supporting Information

**Highly efficient nonvolatile magnetization switching and multi-level states by current in single van der Waals topological ferromagnet $Fe_3GeTe_2$**

*Kaixuan Zhang*, Youjin Lee, Matthew J. Coak, Junghyun Kim, Suhan Son, Inho Hwang, Dong-Su Ko, Youngtek Oh, Insu Jeon, Dohun Kim, Changgan Zeng, Hyun-Woo Lee, and Je-Geun Park**

## Contents:

### Supporting Figures:

## Supporting Figures

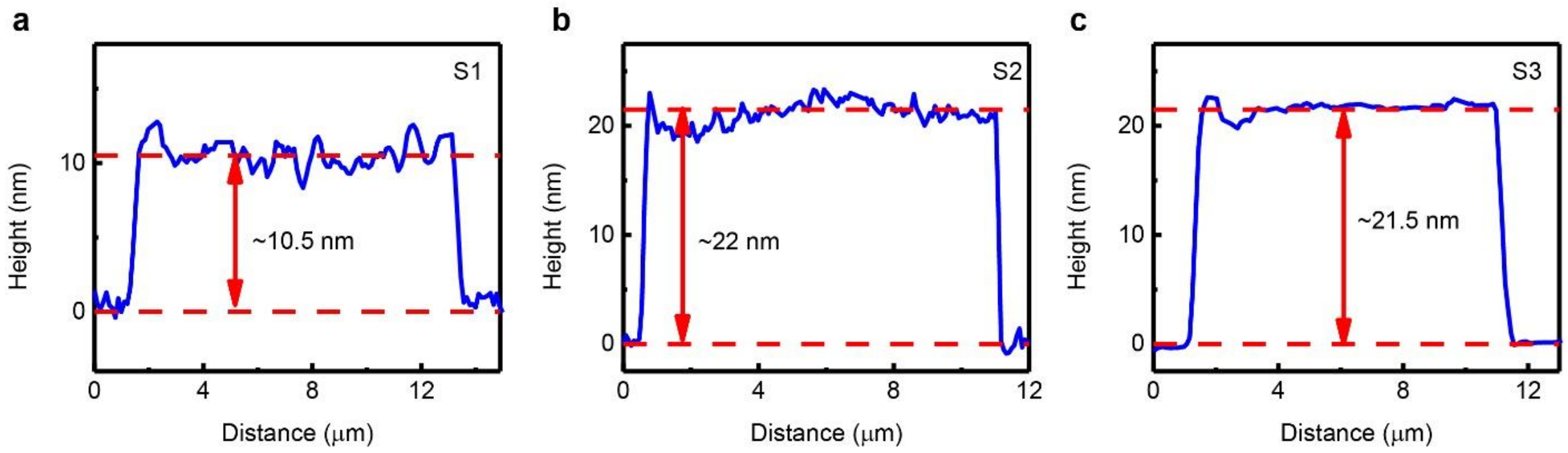

**Figure S1.** a-c) The thickness of samples S1, S2, S3 are ~10.5 nm, ~22 nm, ~21.5 nm, measured by the AFM, respectively.

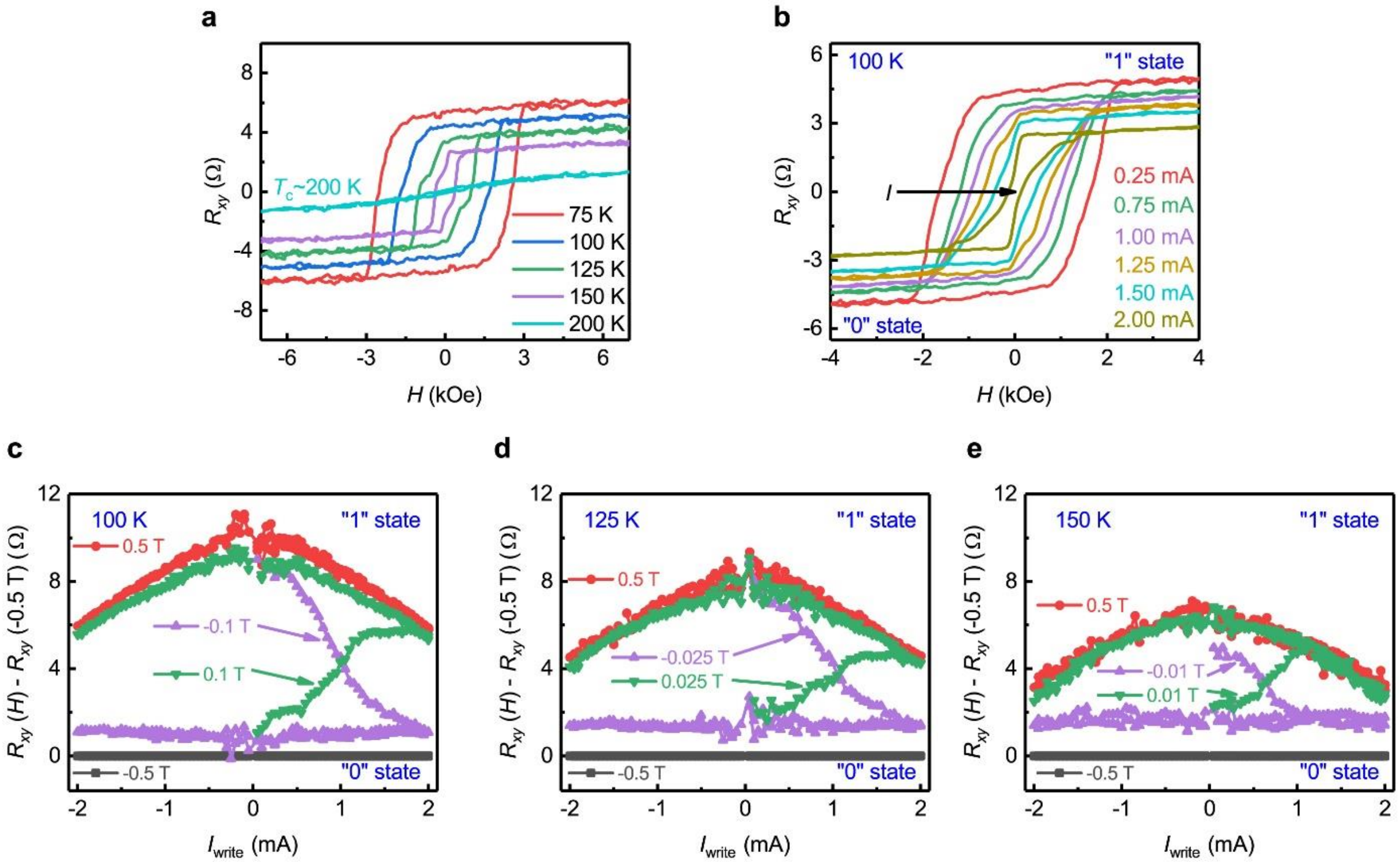

**Figure S2.** The current-controlled switching behavior at higher temperatures above the liquid nitrogen temperature. (a) $R_{xy}$-$H$ curves at various temperatures. (b) $R_{xy}$-$H$ curves at 100 K with various current $I$. (c-e) The current-controlled switching behavior at 100, 125, and 150 K, respectively. $R_{xy}$ ($H$)-$R_{xy}$ (-0.5 T) as a function of $I_{write}$ (0 → 2 → -2 → 2 mA) under various magnetic fields from -0.5 T to 0.5 T. The green and purple arrows indicate the switching from "0" to "1" and "1" to"0" states while ramping $I_{write}$ from 0 to 2 mA, respectively.

.

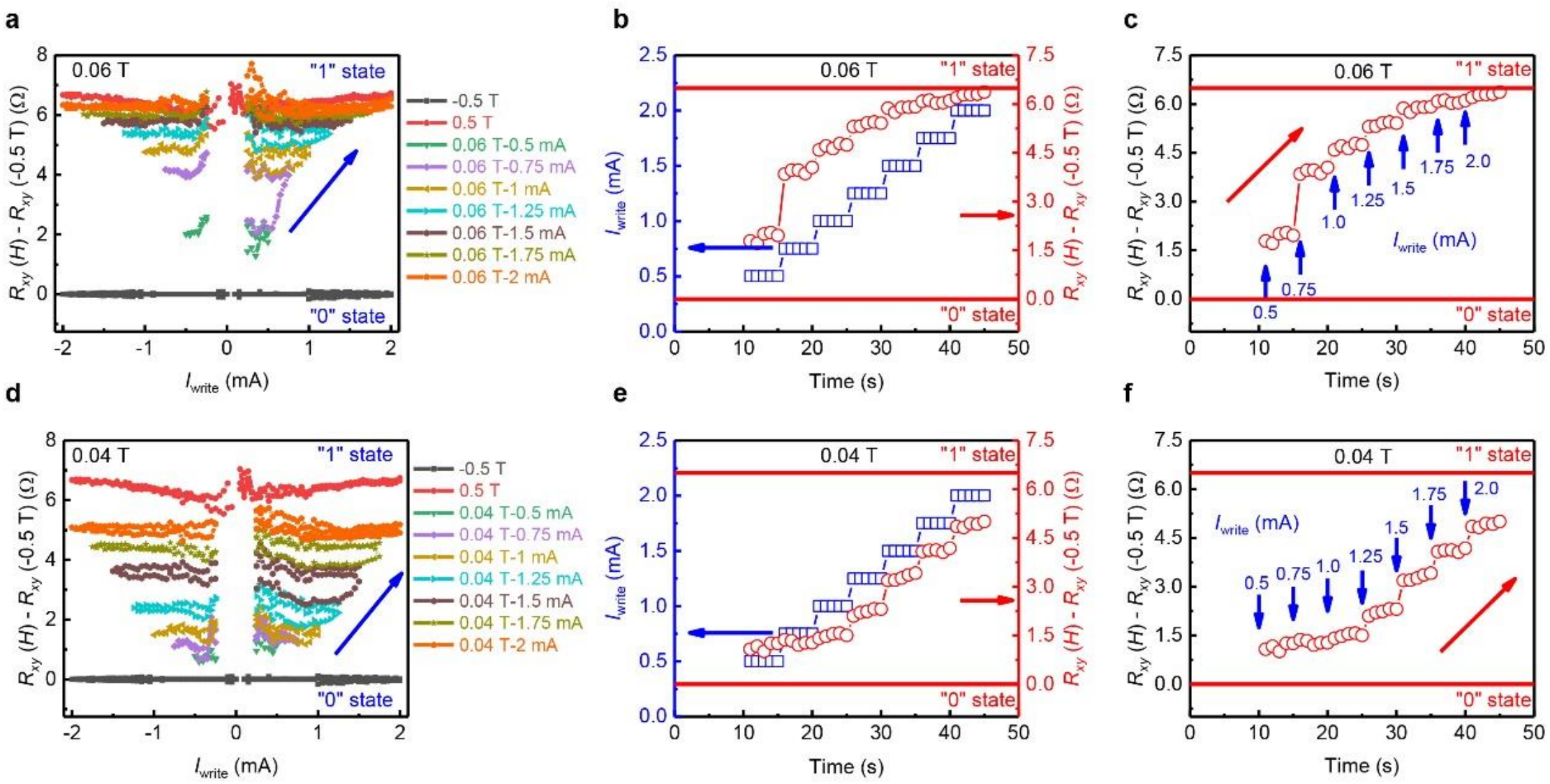

**Figure S3.** Multi-level states are achieved through the application of current with systematic performance. a) $R_{xy}$ ($H$)-$R_{xy}$ (-0.5 T) as a function of $I_{write}$ (0 → 0.5 → -0.5 → 0.5 → 0 mA; then 0 → 0.75 → -0.75 → 0.75 → 0 mA) under 0.06 T. The system gradually enters into different intermediate states from "0" to "1" states (indicated by the blue arrow). b) $R_{xy}$ ($H$)-$R_{xy}$ (-0.5 T) as a function of time with different maximum $I_{write}$. The blue curve represents the time sequence of $I_{write}$, while the red curve denotes the magnetization's corresponding time sequence. Multiple step-like states were achieved from "0" to "1" state. c) $R_{xy}$ ($H$)-$R_{xy}$ (-0.5 T) as a function of time with different maximum $I_{write}$. The blue arrows denote the value of applied $I_{write}$ as time elapses. Multiple step-like states were achieved from the "0" to "1" states (indicated by the red arrow). d-f) The same behavior but under a magnetic field of 0.04 T.

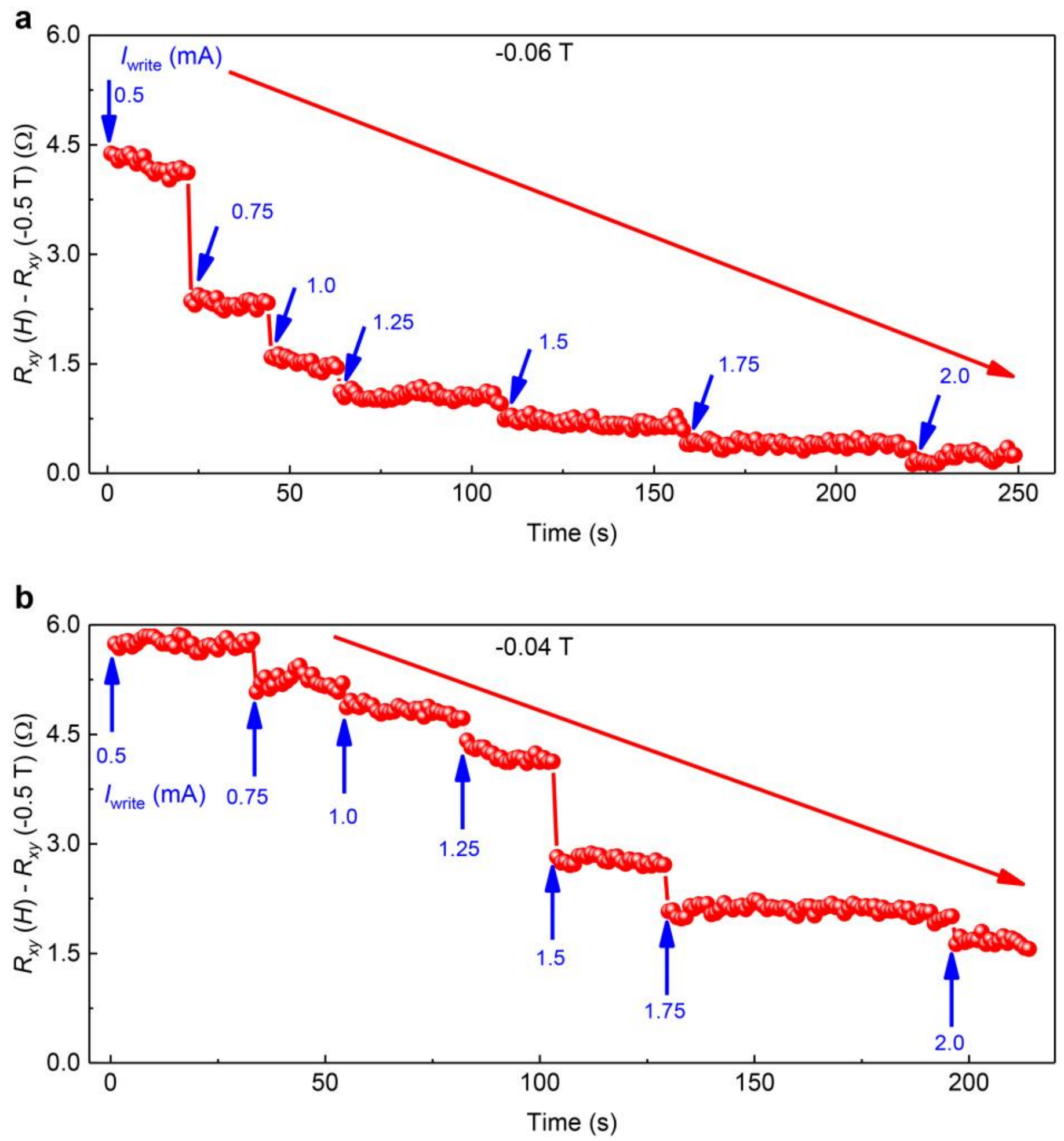

**Figure S4.** The current-controlled multi-level states versus time. $R_{xy}$ ($H$)-$R_{xy}$ (-0.5 T) as a function of time with different $I_{write}$ under a magnetic field of -0.06 T (a) and -0.04 T (b), respectively.